\documentclass[aps,prl,twocolumn, groupedaddress]{revtex4-1}
\usepackage{graphicx}
\begin{document}

%\preprint{}

\title{Strong electrically tunable exciton g-factors in individual quantum dots due to hole orbital angular momentum quenching.}


\author{V. Jovanov}
\email[]{jovanov@wsi.tum.de}
\author{T. Eissfeller}
\author{S. Kapfinger}
\author{E. C. Clark}
\author{F. Klotz}
\author{G. Abstreiter}
\author{J. J. Finley}
%\thanks{}
\affiliation{Walter Schottky Institut, Technische Universit\"{a}t M\"{u}nchen, Am Coulombwall 3, 85748 Garching, Germany}


\date{\today}

\begin{abstract}
Strong electrically tunable exciton g-factors are observed in individual (Ga)InAs self-assembled quantum dots and the microscopic origin of the effect is explained. Realistic eight band $\textbf{k}\cdot\textbf{p}$ simulations quantitatively account for our observations, simultaneously reproducing the exciton transition energy, DC Stark shift, diamagnetic shift and g-factor tunability for model dots with the measured size and a comparatively low In-composition of $x_{In}\sim35\%$ near the dot apex. We show that the observed g-factor tunability is dominated by the \textit{hole}, the electron contributing only weakly. The electric field induced perturbation of the \textit{hole} wavefunction is shown to impact upon the g-factor via orbital angular momentum quenching, the change of the In:Ga composition inside the envelope function playing only a minor role. Our results provide design rules for growing self-assembled quantum dots for electrical spin manipulation via electrical g-factor modulation. 

\end{abstract}

%\keywords{}

\maketitle

%----------------------------------------------------------------------------------------------------------------------------------------
% INDRODUCTION AND MOTIVATION
% JJF last change 19th August 2010 
%----------------------------------------------------------------------------------------------------------------------------------------
The spin of charge carriers in semiconductor quantum dots (QDs) has recently attracted much attention due to the promise it may provide for solid-state quantum information processing \cite{Hanson2007,Krenner2010}. In this respect, the need to selectively rotate a \textit{specific} spin qubit within a quantum register, whilst simultaneously controlling interactions between spins is a challenging task. Such selective addressing either requires that each qubit has a unique resonance frequency or calls for highly local ($\leq100$nm) time dependent magnetic fields to selectively rotate a specific qubit \cite{Koppens2006,Kroner2008}. Methods to create nanoscale time dependent magnetic fields do not exist, motivating recent proposals for electrical spin control in both single layer \cite{Pingenot2008} QDs and QD-molecules \cite{Andlauer2009} via Land$\acute{e}$ g-tensor modulation. Such approaches aim to tune the magnetic response using electric fields to push the carrier envelope functions into different regions of the nanostructure. This provides the potential to achieve arbitrary spin rotations on the Bloch sphere by applying time dependent \textit{electric} fields \cite{Pingenot2008,Andlauer2009} using e.g. metallic gate structures. To date, electrical g-factor modulation in semiconductor nanostructures has been demonstrated using parabolically composition graded AlGaAs quantum wells \cite{Salis2001} and vertically coupled (Ga)InAs QD-molecules \cite{Doty2006}. However, very weak effects are typically observed for single dots \cite{Nakaoka2007}. Recently, we reported electrically tunable exciton g-factors in (Ga)InAs self-assembled QDs grown using the partially covered island (PCI) method but could not identify the mechanism responsible for the tuning \cite{Klotz2010}.\\     
%----------------------------------------------------------------------------------------------------------------------------------------      
In this letter, we observe very strong electrical tunability of the exciton g-factor ($g_{ex}=g_e+g_h$) in (Ga)InAs self-assembled QDs grown \textit{without} the PCI method and unambiguously identify its microscopic origin. By performing realistic eight band $\textbf{k}\cdot\textbf{p}$ simulations using a QD size, shape and In-composition determined by scanning tunneling microscopy, we quantitatively account for experimental results and obtain new insight into the origin of the effect. Our experimental and theoretical findings are in excellent agreement; exciton transition energy, DC Stark shift, diamagnetic shift and g-factor tunability all being \textit{simultaneously} reproduced by theory using dots with a diameter $D=25nm$, height $d=6nm$ and a maximum In-composition of $x_{In}\sim35\%$ near the dot apex. We show that the $g_{ex}$ tunability is dominated by the \textit{hole} ($g_h$), the electron ($g_e$) contributing only weakly. Most surprisingly, the electric field induced perturbation of the hole envelope wavefunction is shown to impact upon $g_h$ principally via orbital angular momentum quenching \cite{pryor:2006}, the change of the In:Ga composition inside the envelope function playing only a minor role. The results show that the strength of the electrical tunability \textit{increases} as the In-alloy content at the dot apex ($x_{In}^{apex}$) \textit{reduces}, explaining why we observe strong electrical $g_{ex}$ tunability in our QDs.

\begin{figure}[htbp]
\includegraphics[width=\columnwidth]{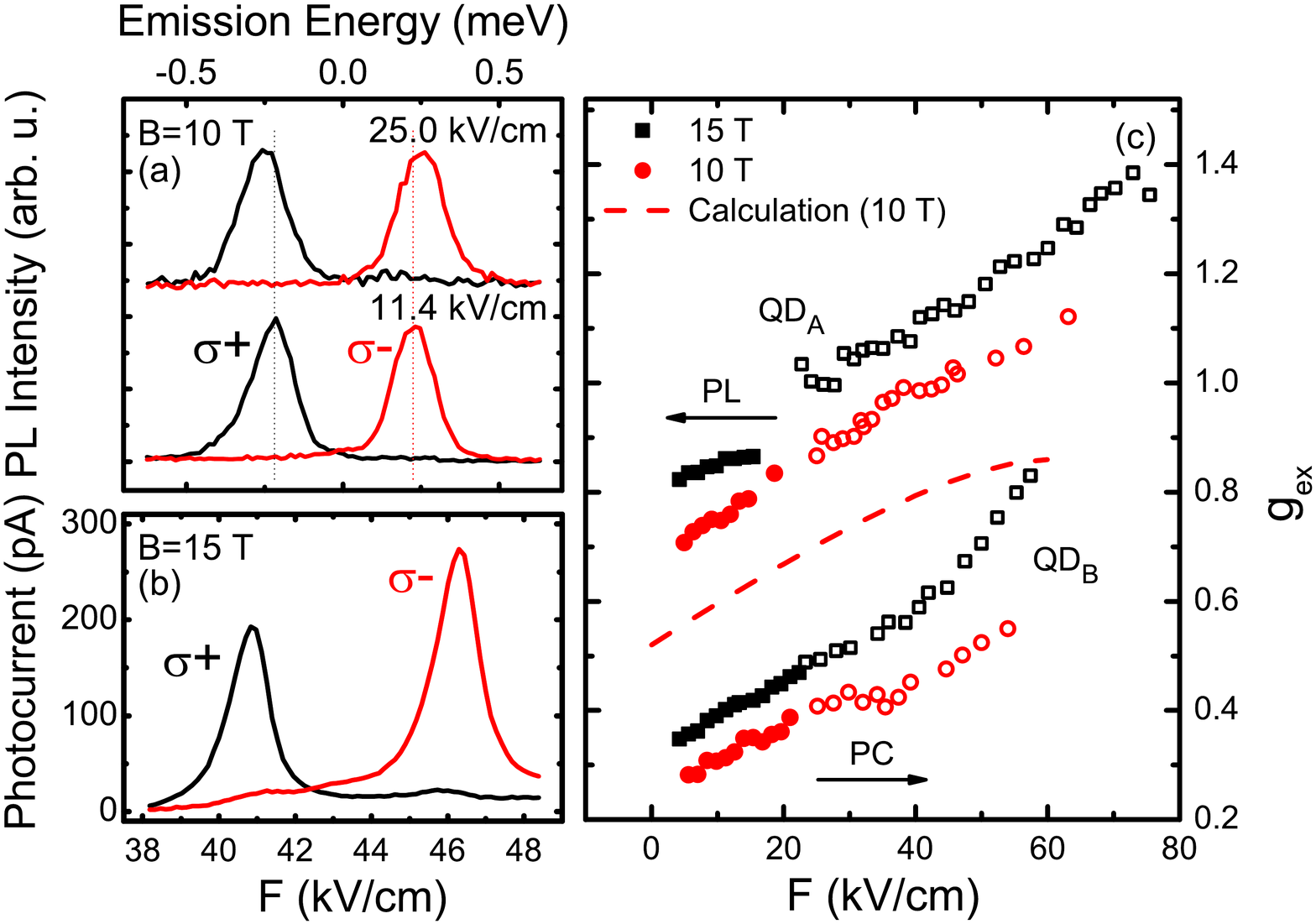}
\caption{\label{fig1} (color online). (\textit{a}) Polarization resolved photoluminescence spectra at two different electric fields and magnetic field of $10$~T; both applied in growth direction. To facilitate a direct comparison of the Zeeman gap ($E_Z=g_{ex}\mu_{B}B$) the Stark shift has been suppressed. (\textit{b}) Polarization resolved photocurrent spectra at magnetic field of $15$~T applied in growth direction. (\textit{c}) Extracted excitonic g-factor as a function of the applied electric field for two representative QDs recorded at $10$~T (circles) and $15$~T (squares). The dashed line shows the results of 8-band $\textbf{k}\cdot\textbf{p}$ calculations using the best fit dot size and composition parameters described in the manuscript.}
\end{figure}

%----------------------------------------------------------------------------------------------------------------------------------------
% DOTS AND SAMPLE DESIGN
% JJF last change 17th August 2010 
%----------------------------------------------------------------------------------------------------------------------------------------
The samples investigated were GaAs n-i-Schottky photodiode structures grown by molecular beam epitaxy. A single layer of nominally In$_{0.5}$Ga$_{0.5}$As self-assembled QDs was grown in the $i$-region at an unusually high growth temperature of $595^{\circ}$C. This is expected to lead to an average In-content significantly lower than the nominal value of $x_{In}=50\%$, due to the combined effects of In-desorption \cite{Heyn2003a} and inter-diffusion with the GaAs matrix material during capping \cite{Heyn2003b}.  Cross sectional STM measurements \cite{keizer:2010} indicate that the resulting dots have a lateral size of $D=26\pm8$nm and height of $h=6\pm2$nm.   
Single dots were optically probed using a low temperature ($4.2$~K) magneto-confocal microscopy set up that facilitates application of magnetic ($B$) fields up to $B=15$~T in Faraday geometry. Typical PL spectra recorded at $B=10$~T and axial electric fields of $11.4$~kV/cm and $25$~kV/cm, respectively, are presented in fig.\ref{fig1}a. The results clearly show two Zeeman split bright excitons with a $\geq95\%$ degree of circular polarization. Upon raising the electric field we observe a clear increase of $g_{ex}$. For electric fields $\geq27$~kV/cm the PL quenches due to carrier tunnelling from the dot and we probe the energy of the Zeeman branches using polarization selective photocurrent (PC) absorption measurements. PC spectra recorded with $\sigma^+$ and $\sigma^-$ polarized excitation were obtained at fixed laser frequency, whilst the levels were tuned into resonance via the DC-Stark effect.  Typical results are plotted in fig.\ref{fig1}b. The dependence of $g_{ex}$ on the electric field is summarized in fig.\ref{fig1}c for two representative QDs labelled $QD_A$ and $QD_B$. For both $QD_A$, $QD_B$ and all other dots investigated $g_{ex}$ increases with the axial electric field.\\
 
%In order to explain this behavior of the excitonic g-factor we developed a theoretical model in which the local In-concentration and In-Ga composition profile is found to strongly influence the orbital angular momentum and the corresponding g-factor of the hole, whilst the electron g-factor remains largely unaffected. The dashed curve presented in the right panel of fig.\ref{fig1} shows the calculated electric field variation of $g_{ex}$ and dot structural parameters presented below.

%----------------------------------------------------------------------------------------------------------------------------------------

% --------------------------------------------------------------------
% THEORY
% Thomas last change 6/15/10
% JJF last change 17th August 2010 
% --------------------------------------------------------------------
To understand our results we performed electronic structure calculations using the eight band $\textbf{k}\cdot\textbf{p}$ envelope function approximation. The magnetic field was introduced into the discrete Hamiltonian in a manifestly gauge invariant manner \cite{andlauer:2008,wilson:1974} and spatial finite volume discretization, combined with the correct operator ordering accounted for abrupt material boundaries \cite{burt:1999}. Strain fields were included using continuum elasticity theory and their impact on the electronic structure was taken into account via deformation potentials and the linear piezoelectric effect \cite{stier:1999}. 
%Atomistic asymmetry and interface effects were \textit{not} included, but have been shown to only introduce weak corrections for large InGaAs quantum dots with low In-content \cite{bester:2005}. Furthermore, the electron-hole exchange interaction active for the $X^0$ bright exciton states can be safely neglected since the mixing of conduction and valence bands is weak for the lowest energy electron and hole orbital states.
The direct Coulomb interaction was included in our calculations using lowest order perturbation theory and the validity of this approximation was carefully checked for a few selected cases where direct electron-hole Coulomb interaction was taken into account in a fully self consistent manner. To obtain quantitative results for the $X^0$ energy and g-factor, a Luttinger-like eight band $\textbf{k}\cdot\textbf{p}$-model was employed, where far-band contributions to the effective mass Hamiltonian and g-factors are included up to the order of $k^2$ \cite{trebin:1979}. We modeled our QDs as having a truncated lens shape with a diameter varying from $D=15-50$~nm, a height of $6$~nm above the wetting layer and an inverse trumpet-like In-compositional profile \cite{offermans:2005, migliorato:2002}. The In-concentration of the InGaAs alloy was taken to be $x_{In}=0.2$ at the base and side of the dot increasing to $x_{In}^{apex}=0.2-0.9$ at the dot apex \cite{migliorato:2002}. This range of parameters are fully consistent with the results of cross sectional STM measurements performed on samples grown under the same conditions, from which we also determined the wetting layer thickness ($2$~nm) and $x_{In}^{WL}=0.18$ \cite{keizer:2010}.\\

%THIS SEEMS TO INTERRUPT THE FLOW, IS A DETAIL AND COULD GO IN AN ENDNOTE  
%Note that the sign convention of $g_{ex}$ in this work is as according to:\begin{equation} g_e=-g^n \textrm{, } g_h = g^n \textrm{ and } g_x = g_e + g_h \textrm{,} \end{equation} where $g^n\mu_{B}B=E^{n\uparrow}-E^{n\downarrow}$ is the g-factor of the $n^{th}$ electronic state in the purely electronic picture where the positive $[001]$ direction was chosen to be the quantization axis $z$ of the electron spin with the usual basis of Pauli matrices such that spin up (spin down) is eigenvector of $\sigma_z$ correspong to the eigenvalue 1 (-1). This choice of the quantization axis for the electronic spin is natural in multi-band calculations for a B-field applied along the $z$-direction.  However, we note that this is not necessarily the case in single-band calculations and in experiments where the circular polarization direction may be defined as \textit{seen by the sample} and \textit{as seen from the laser/detector} as well. This is one reason that conflicting notations for the \textit{sign} of the electron ($g_e$), hole ($g_h$) and exciton ($g_{ex}$) g-factors can be found in the literature.

\begin{figure}[htbp]
\includegraphics[width=\columnwidth]{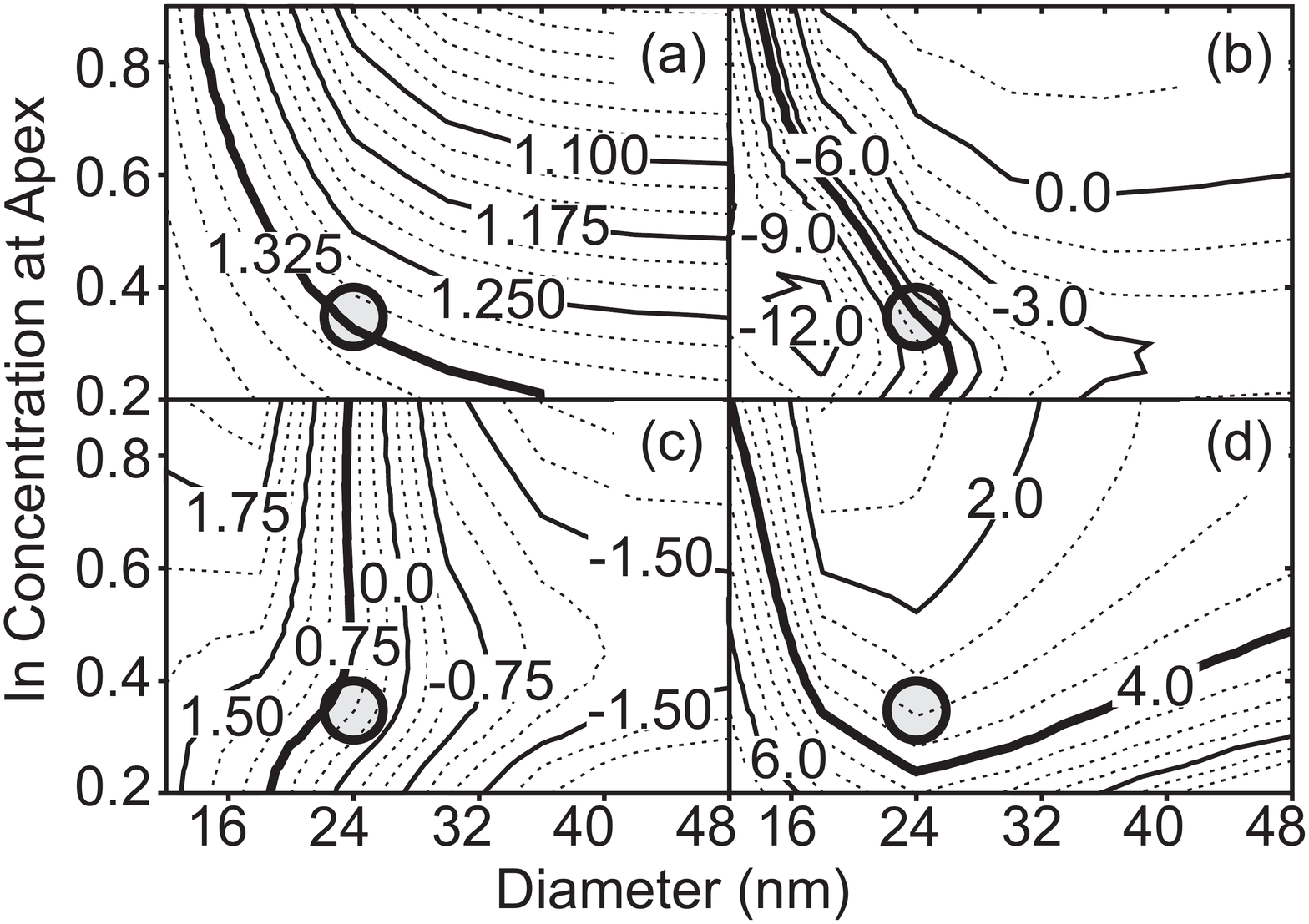}
\caption{\label{fig:2} The following properties were obtained by varying QD diameter and InAs content in our calculations: (\textit{a}) The $X^0$ exciton energy (eV) at $7\,\textrm{kV}/\textrm{cm}$ and $10\,\textrm{T}$, (\textit{b}) the Stark shift energy (meV) between $0$ and $60\,\textrm{kV}/\textrm{cm}$ at $10\,\textrm{T}$, (\textit{c}) the exciton g-factor at zero electric field and $10\,\textrm{T}$, and (\textit{d}) the diamagnetic shift energy (meV) between $1$ and $13\,\textrm{T}$ at $7\,\textrm{kV}/\textrm{cm}$. The circle indicates the QD parameters used in other calculations and the bold lines indicate experimentally measured values.}
\end{figure}

We continue to discuss the results of our calculations: Figures \ref{fig:2}a, \ref{fig:2}b and \ref{fig:2}c show contour plots of the exciton transition energy at $7$~kV/cm, Stark shift from $0-60$~kV/cm and $g_{ex}$ at $B=10$~T, respectively.  Figure \ref{fig:2}d shows the diamagnetic shift from $1$~T to $13$~T at an electric field of $7$~kV/cm. The measured values of these quantities for $QD_A$ are represented by the bold contours on the figure, showing that all are reproduced in the $D$,$x_{In}^{apex}$ parameter space probed.  More importantly, all of these contours intersect at $D=24\pm2$~nm and $x_{In}^{apex}=0.35\pm0.02$, as indicated by the open circles on the various panels of fig.\ref{fig:2}. We note that we also calculated very different $X^0$ g-factors for small ($D\leq15$~nm), InAs rich ($x_{In}^{apex}$) dots, in very good agreement with previous experiments and calculations \cite{nakaoka:2004}. %the quantization direction of spin appears to be reversed in this ref.
%From the observed diamagnetic and Stark shift contours we can clearly rule out much larger dots having the same $X^0$ energy but much lower In-content.\\

\begin{figure}[htbp]
\includegraphics[width=\columnwidth]{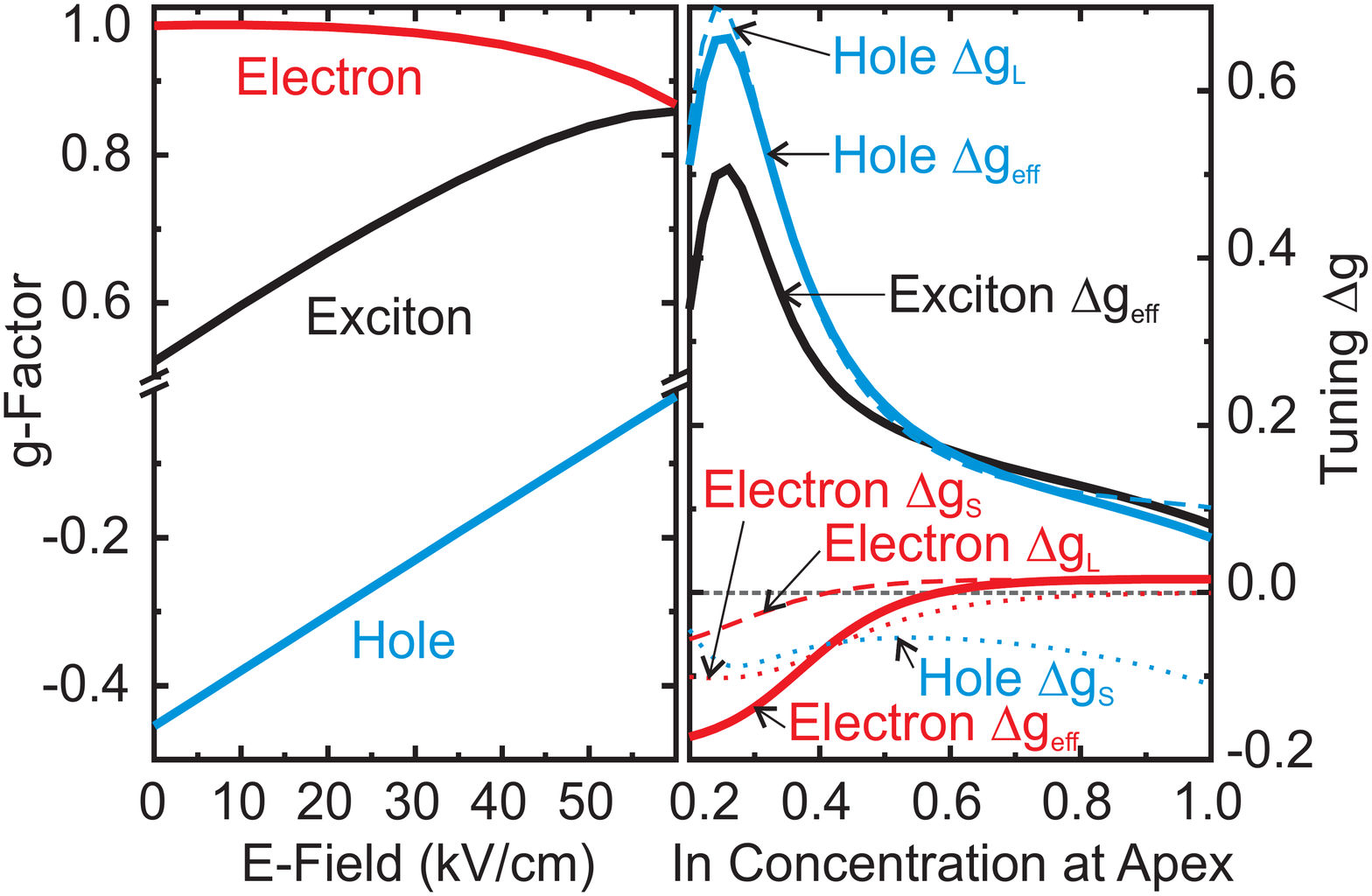}
\caption{\label{fig:3} (\textit{left}) Calculated electron, hole and exciton g-factor as a function of the electric field. (\textit{right}) Calculated dependency of the electric tuning of $g_e$, $g_h$ and $g_{ex}$ as a function of In-concentration at the dot apex. The electric tuning is defined as g-factor at $60\,\textrm{kV}/\textrm{cm}$ minus g-factor at $0\,\textrm{kV}/\textrm{cm}$.}
\end{figure}

The electron and hole g-factors in self-assembled dots have been investigated in previous theoretical works; primary contributions arising from (i) strain induced band-mixing \cite{nakaoka:2005}, (ii) modification of Roth's formula by the effective band-gap \cite{roth:1959, pryor:2006} and (iii) orbital angular momentum quenching \cite{kiselev:1998, lommer:1985, pryor:2006}. Electric field induced changes in the alloy overlap, i.e. the In-Ga content within the envelope function, have been shown to be important mostly for very extended electronic states in weakly confined dots \cite{pryor:2006, nakaoka:2004}. However, whilst each of the effects (i)-(iii) have been reported to contribute to $g_{ex}$, the microscopic origin of the strong electrical $g_{ex}$ tunability in our samples is not at all obvious. We now demonstrate that the observed effects can be directly traced to strong electric field induced changes of the \textit{hole} g-factor ($g_h$), the electron g-factor ($g_e$) being much more weakly influenced by the electric field. The left panel of fig.\ref{fig:3} shows representative calculations of $g_h$, $g_e$ and $g_{ex}=g_h+g_e$ for $B=10$~T using the model dot parameters deduced from fig.\ref{fig:2} ($D=24$~nm and $x_{In}^{apex}=0.35$). Clearly $g_e$ varies only weakly over the range of electric fields investigated ($\Delta g_{e}/g_{e}\leq10\%$), whilst $g_h$ is much more strongly affected ($\Delta g_{h}/g_{h}\geq50\%$). 

We now show that quenching of the orbital angular momentum in zero dimensional structures (mechanism - (iii)) is primarily responsible for the observed electrical tunability. The contributions to the g-factor of any electronic state with orbital index $n$ and spin orientation $\uparrow$ or $\downarrow$ ( $|n,\uparrow\rangle$ and $|n,\downarrow\rangle$) can be written as \cite{kiselev:1998}
\begin{equation}
g_n^{eff}=g_0 + g_{L\,n} + g_{S\,n},
\label{eq:geff}
\end{equation}
where $g_0\approx 2$ is the free electron Land$\acute{e}$ g-factor and $g_{S\,n}$ is the contribution of remote bands included perturbatively in the 8-band $\textbf{k}\cdot\textbf{p}$ model, a function of the In-content within the envelope function. In comparison, $g_{L\,n}$ is the contribution to the g-factor due to the angular motion of the electron. This quantity can be obtained using second order perturbation theory and has the form,
% I changes the spacings by hand to handle the overflow...
\begin{equation}
g_{L\,n}\hspace{-1mm}=\hspace{-1mm}-\frac{2 m_0}{\hbar^2} 
\hspace{-4mm}
\sum_{
\begin{array}{c}
n' \neq n\\
s'
\end{array}}
\hspace{-4mm}
{\frac{\left| \left< n\uparrow \right|\hat{P}_+\left|n'\,s'\right> \right|^2 \hspace{-2mm} - \hspace{-1mm} \left| \left< n\uparrow \right|\hat{P}_-\left|n'\,s'\right> \right|^2 }
{E_n - E_{n'}}}
\textrm{,}
\hspace{-1mm}
\label{eq:gl_2nd}
\end{equation}
with a momentum operator $\hat{P}_\pm = \hat{P}_x \pm i\hat{P}_y$. The sum runs over all conduction and valence band states $\left|n'\,s'\right>$ except $n' = n$.
Here, $\mathbf{\hat{P}}=\frac{1}{\hbar}\partial_{\mathbf{k}}\hat{H}(\mathbf{k})$ is the 8-band momentum operator obtained by the Helman-Feynman theorem \cite{foreman:2000} and the quantization axis $z$ is taken to be $\left\langle001\right\rangle$, the direction of the applied static magnetic field. In a fully equivalent manner \cite{kiselev:1998}, we can also express $g_{L\,n}$ in the framework of \textsl{first} order perturbation theory

\begin{equation}
g_{L\,n}=-\left(\left<n \uparrow \right|\hat{L}_z\left| n \uparrow\right> - \left< n\downarrow\right|\hat{L}_z\left|n \downarrow\right> \right)
\textrm{,}
\label{eq:gl_1nd}
\end{equation}

with the orbital angular momentum operator $\hat{L}_z = \left( \mathbf{\hat{r}} \times \mathbf{\hat{P}} \right)_z$.
% The following explains how one can (approximately) obtain eqn. \ref{eq:gl_1nd} from simple arguments.
% Helpful, but not necessary
%In order to provide more insight in the physical meaning of eqn. \ref{eq:gl_1nd} we will qualitatively derive this equation:
%Observing that in the centered symmetric gauge the vector potential $\mathbf{A}$ can be written as $\mathbf{A} = \frac{1}{2} \mathbf{\hat{r}} \times \mathbf{B}$ and $\mathbf{B} = (0,0,B_z)^T$, the coupling of the electron motion to the vector potential in Schr"odinger's Equation is $\frac{e \hbar}{2 m_0}\mathbf{A}.\mathbf{P} = \mu_B (\mathbf{\hat{r}} \times \mathbf{B}).\mathbf{P}=-\frac{1}{2} \mu_B B_z (\mathbf{\hat{r}} \times \mathbf{\hat{P}})_z = -\frac{1}{2} \mu_B B_z \hat{L}_z = \frac{1}{2} B_z \mu_L$ with the effective mangetic moment $\mu_L = -\mu_B L_z$ of the electron motion and the Bohr magneton $\mu_B$. Then, with $g_L = 2 \mu_L / \mu_B$ and $\left<n \uparrow \right|\hat{L}_z\left| n \uparrow\right> \approx - \left< n\downarrow\right|\hat{L}_z\left|n \downarrow\right>$ we obtain eqn \ref{eq:gl_1nd}. This approximation can be obtained formally by fixing the gauge (therby breaking gauge invariance) and performing the Peiler's substitution. This has been shown to be reasonable in the discretized Hamiltonian only for the lowest electron and hole states \cite{governal:1998}.
Eqn. \ref{eq:gl_1nd} can be readily evaluated to obtain the tunability of the angular momentum g-factor $\Delta g_L$. The right panel of fig.\ref{fig:3} compares the calculated value of $\Delta g_{e/h}=g_{e/h}(60kV/cm)-g_{e/h}(0kV/cm)$ obtained from our full calculation (solid lines, labeled $\Delta g_{eff}$) and the contributions from $\Delta g_L$ obtained using eqn. \ref{eq:gl_1nd} (dashed lines) and the far band contribution $\Delta g_S$ (dotted lines) for $0.2 \leq x_{In}^{apex} \leq 1.0$.  

We find several prominent features of which we highlight two: Firstly, the electric field tuning of $g_h$ is almost entirely due to the angular momentum contribution.  In comparison, $g_L$ plays a much less important role for the electrical modification of $g_e$. The electrically induced modification of the far-band correction ($g_S$) is weak for both hole and electron. Secondly, there is a clear maximum in the tunability of $g_h$ for \textit{low} In-concentrations at the dot apex ($0.25 \leq x_{In}^{apex} \leq 0.35$). Thus, we conclude that the observed electrical tunability stems from the modification of the orbital angular momentum of the valence band state and that the low In-content of the presently studied QDs increases the strength of the effect.  
The increase of the angular momentum contribution $g_L$ to the hole ground state is equivalent to a decrease of angular momentum according to eqn. \ref{eq:gl_1nd}. 
Whilst we evaluated eqn. \ref{eq:gl_1nd} to obtain the $\Delta g_L$ curves presented in the right panel of fig.\ref{fig:3}, we now make use of eqn. \ref{eq:gl_2nd} to obtain a qualitative understanding of the observed $g_h$ tunability: The conduction band (CB) states in a III-V dot have large momentum matrix elements (MMEs), defined by $\left| \left< n\uparrow \right|\hat{P}_{\pm}\left|n'\,s'\right> \right|^2$ in eqn. \ref{eq:gl_1nd}, with the valence band (VB) states. In particular, the MME between the lowest energy electron and hole orbital states without electric field is large. For In-dilute dots there are many bound hole states but only a few bound electron states. For example, we calculate that the best-fit QD from fig.\ref{fig:2} with $D=24nm$ and $x_{In}^{apex}=0.35$ accommodates $>24$ bound hole states, but only 3 bound electron states. The electric field decreases the MME between specific pairs of bound CB and VB orbital states (e.g. lowest orbital states) whilst this modification is compensated by an increase of the MME between other orbital states according to the \textit{f}-sum rule. For the electron ground state, there are \textit{many} bound hole states available to compensate for the field induced reduction of the MME between the lowest CB and VB orbital.  Each of these VB orbital states has approximately the same relative energy compared to the effective band gap of dot ($\sim1300$~meV).  Thus, the energy denominator of eqn. \ref{eq:gl_2nd} is approximately the same for each pair of CB and VB states and there is almost no change in the angular momentum of the electron state and a weak tunability of $g_e$. In strong contrast, for the lowest energy VB state there are very \textit{few} bound CB states that can compensate for the field induced change of the MME.  Thus, $g_h$ is strongly influenced by the electric field and dominates the observed tunability of $g_{ex}$.  
%Furthermore, for high fields (60 kV/cm) only one bound electron state remains in dot. 
%The wetting layer electron states have virtually no overlap with the hole localized at the top of the dot. Thus, the states that compensate changes of the momentum matrix elements are the bulk GaAs continuum with higher relative energies. 
%Electron states in the wetting layer have no overlap with the hole localized at the top of the dot. States in the GaAs continuum have higher relative energies and therefore a larger energy denominator in eqn. \ref{eq:gl_2nd} that quenches the angular momentum and increases the g-factor.
%This quenching can be seen directly in fig.~\ref{fig:3} (\textit{right}) where the calculated density of the LH$\uparrow$ component ($\approx 5\%$) of the spin $\uparrow$ hole is shown as seen from the top. 
%The electric field compresses this component and reduces the angular momentum since $\hat{L}_z$ is proportional to the radius vector.
%The sketch in fig.~\ref{fig:3} illustrates the lateral compression of the hole state and the reduction of electron-hole overlap induced by the field. 

It has been speculated that the change in alloy overlap may induce HH-LH mixing that could be responsible for the observed strong electrical tunability \cite{Klotz2010}, since the HH and LH g-factors differ strongly. However, our calculations show that ramping the electric field from $0-60$kV/cm results in only a weak change of the LH admixture of the lowest VB orbital state ($\approx 0.2\%$). %It has been speculated that the electric field induces a change in the alloy overlap of the wave functions.
Similarly, the electric field leads to no substantial change of the In:Ga alloy content within the hole envelope function. We calculated that for $x_{In}^{apex}=0.35$ and $D=24$ nm, the In-alloy overlap changes by $\approx 0.14\%$ or $\approx -2.2\%$ for the hole and electron, respectively. This leads to an extremely weak change in the far-band correction $\Delta g_S$ as shown in the right panel of fig.\ref{fig:3} and does not significantly contribute to the observed electric field tunability. %(\textit{bottom})) and in the LH component of the hole ($\approx +0.002$).
In summary, we identified the microscopic origin of pronounced electrical tunability of the exciton g-factor in composition engineered (Ga)InAs self-assembled QDs.
%By comparing our results with eight band $\textbf{k}\cdot\textbf{p}$ simulations using a QD size, shape and In-composition determined from structural microscopy, we obtained excellent consistent agreement with several observables for dots with a diameter $D=24nm$, height $h=6nm$ and a maximum In-composition of $x_{In}\sim35\%$ near the dot apex.
We demonstrated that the $g_{ex}$ tunability is dominated by $g_h$, $g_e$ contributing only weakly. The electric field induced perturbation of the hole envelope wavefunction was shown to impact upon $g_h$ principally via orbital angular momentum quenching, the change of the In:Ga composition inside the envelope function playing only a minor role. %Finally, the strength of the electrical tunability was shown to be strong only for dots with a low In-alloy content at the dot apex.
Our results provide significant scope for morphological and structural tailoring self-assembled QDs to allow all electrical spin control via the g-tensor \cite{Pingenot2008}. 

\begin{acknowledgments}
This work is funded by the DFG via SFB-631 and NIM and the EU via SOLID. We gratefully acknowledge J. Keizer and P. Koenraad for useful discussions and for performing the X-STM measurements on our samples.
\end{acknowledgments}

\bibliography{g-fac_bib}

%Merlin.mbs v4.21 2009-07-09.
\providecommand{\noopsort}[1]{}\providecommand{\singleletter}[1]{#1}%
\begin{thebibliography}{10}%
\makeatletter
\providecommand \@ifxundefined [1]{%
 \ifx #1\undefined \expandafter \@firstoftwo
 \else \expandafter \@secondoftwo
\fi
}%
\providecommand \@ifnum [1]{%
 \ifnum #1\expandafter \@firstoftwo
 \else \expandafter \@secondoftwo
\fi
}%
\providecommand \enquote [1]{``#1''}%
\providecommand \bibnamefont  [1]{#1}%
\providecommand \bibfnamefont [1]{#1}%
\providecommand \citenamefont [1]{#1}%
\providecommand\href[0]{\@sanitize\@href}%
\providecommand\@href[1]{\endgroup\@@startlink{#1}\endgroup\@@href}%
\providecommand\@@href[1]{#1\@@endlink}%
\providecommand \@sanitize [0]{\begingroup\catcode`\&12\catcode`\#12\relax}%
\@ifxundefined \pdfoutput {\@firstoftwo}{%
 \@ifnum{\z@=\pdfoutput}{\@firstoftwo}{\@secondoftwo}%
}{%
 \providecommand\@@startlink[1]{\leavevmode\special{html:<a href="#1">}}%
 \providecommand\@@endlink[0]{\special{html:</a>}}%
}{%
 \providecommand\@@startlink[1]{%
  \leavevmode
  \pdfstartlink
   attr{/Border[0 0 1 ]/H/I/C[0 1 1]}%
   user{/Subtype/Link/A<</Type/Action/S/URI/URI(#1)>>}%
  \relax
 }%
 \providecommand\@@endlink[0]{\pdfendlink}%
}%
\providecommand \url  [0]{\begingroup\@sanitize \@url }%
\providecommand \@url [1]{\endgroup\@href {#1}{\urlprefix}}%
\providecommand \urlprefix [0]{URL }%
\providecommand \Eprint[0]{\href }%
\@ifxundefined \urlstyle {%
  \providecommand \doi [1]{doi:\discretionary{}{}{}#1}%
}{%
  \providecommand \doi [0]{doi:\discretionary{}{}{}\begingroup
  \urlstyle{rm}\Url }%
}%
\providecommand \doibase [0]{http://dx.doi.org/}%
\providecommand \Doi[1]{\href{\doibase#1}}%
\providecommand \bibAnnote [3]{%
  \BibitemShut{#1}%
  \begin{quotation}\noindent
    \textsc{Key:}\ #2\\\textsc{Annotation:}\ #3%
  \end{quotation}%
}%
\providecommand \bibAnnoteFile [2]{%
  \IfFileExists{#2}{\bibAnnote {#1} {#2} {\input{#2}}}{}%
}%
\providecommand \typeout [0]{\immediate \write \m@ne }%
\providecommand \selectlanguage [0]{\@gobble}%
\providecommand \bibinfo [0]{\@secondoftwo}%
\providecommand \bibfield [0]{\@secondoftwo}%
\providecommand \translation [1]{[#1]}%
\providecommand \BibitemOpen[0]{}%
\providecommand \bibitemStop [0]{}%
\providecommand \bibitemNoStop [0]{.\EOS\space}%
\providecommand \EOS [0]{\spacefactor3000\relax}%
\providecommand \BibitemShut [1]{\csname bibitem#1\endcsname}%
%</preamble>
\bibitem{Hanson2007}%
  \BibitemOpen
  \bibfield{author}{%
  \bibinfo {author} {\bibfnamefont{R.}~\bibnamefont{Hanson}}, \bibinfo {author}
  {\bibfnamefont{L.~P.}\ \bibnamefont{Kouwenhoven}}, \bibinfo {author}
  {\bibfnamefont{J.~R.}\ \bibnamefont{Petta}}, \bibinfo {author}
  {\bibfnamefont{S.}~\bibnamefont{Tarucha}},\ and\ \bibinfo {author}
  {\bibfnamefont{L.~M.~K.}\ \bibnamefont{Vandersypen}},\ }%
  \bibfield{journal}{%
  \bibinfo {journal} {Rev. Mod. Phys.}\ }%
  \textbf{\bibinfo {volume} {79}},\ \bibinfo {pages} {1217} (\bibinfo {year}
  {2007})%
  \bibAnnoteFile{NoStop}{Hanson2007}%
\bibitem{Krenner2010}%
  \BibitemOpen
  \bibfield{author}{%
  \bibinfo {author} {\bibfnamefont{H.~J.~K.}\ \bibnamefont{O.~Gywat}}\ and\
  \bibinfo {author} {\bibfnamefont{J.}~\bibnamefont{Berezovsky}},\ }%
  \emph{\bibinfo {title} {Spins in Optically Active Quantum Dots}}\ (\bibinfo
  {publisher} {Wiley-VCH},\ \bibinfo {year} {2010})\ ISBN \bibinfo {isbn}
  {978-3-527-40806-1}%
  \bibAnnoteFile{NoStop}{Krenner2010}%
\bibitem{Koppens2006}%
  \BibitemOpen
  \bibfield{author}{%
  \bibinfo {author} {\bibfnamefont{F.~H.~L.}\ \bibnamefont{Koppens}}, \bibinfo
  {author} {\bibfnamefont{C.}~\bibnamefont{Buizert}}, \bibinfo {author}
  {\bibfnamefont{K.~J.}\ \bibnamefont{Tielrooij}}, \bibinfo {author}
  {\bibfnamefont{I.~T.}\ \bibnamefont{Vink}}, \bibinfo {author}
  {\bibfnamefont{K.~C.}\ \bibnamefont{Nowack}}, \bibinfo {author}
  {\bibfnamefont{T.}~\bibnamefont{Meunier}}, \bibinfo {author}
  {\bibfnamefont{L.~P.}\ \bibnamefont{Kouwenhoven}},\ and\ \bibinfo {author}
  {\bibfnamefont{L.~M.~K.}\ \bibnamefont{Vandersypen}},\ }%
  \bibfield{journal}{%
  \bibinfo {journal} {Nature}\ }%
  \textbf{\bibinfo {volume} {442}},\ \bibinfo {pages} {766} (\bibinfo {year}
  {2006})%
  \bibAnnoteFile{NoStop}{Koppens2006}%
\bibitem{Kroner2008}%
  \BibitemOpen
  \bibfield{author}{%
  \bibinfo {author} {\bibfnamefont{M.}~\bibnamefont{Kroner}}, \bibinfo {author}
  {\bibfnamefont{K.~M.}\ \bibnamefont{Weiss}}, \bibinfo {author}
  {\bibfnamefont{B.}~\bibnamefont{Biedermann}}, \bibinfo {author}
  {\bibfnamefont{S.}~\bibnamefont{Seidl}}, \bibinfo {author}
  {\bibfnamefont{S.}~\bibnamefont{Manus}}, \bibinfo {author}
  {\bibfnamefont{A.~W.}\ \bibnamefont{Holleitner}}, \bibinfo {author}
  {\bibfnamefont{A.}~\bibnamefont{Badolato}}, \bibinfo {author}
  {\bibfnamefont{P.~M.}\ \bibnamefont{Petroff}}, \bibinfo {author}
  {\bibfnamefont{B.~D.}\ \bibnamefont{Gerardot}}, \bibinfo {author}
  {\bibfnamefont{R.~J.}\ \bibnamefont{Warburton}},\ and\ \bibinfo {author}
  {\bibfnamefont{K.}~\bibnamefont{Karrai}},\ }%
  \bibfield{journal}{%
  \bibinfo {journal} {Phys. Rev. Lett.}\ }%
  \textbf{\bibinfo {volume} {100}},\ \bibinfo {pages} {156803} (\bibinfo {year}
  {2008})%
  \bibAnnoteFile{NoStop}{Kroner2008}%
\bibitem{Pingenot2008}%
  \BibitemOpen
  \bibfield{author}{%
  \bibinfo {author} {\bibfnamefont{J.}~\bibnamefont{Pingenot}}, \bibinfo
  {author} {\bibfnamefont{C.~E.}\ \bibnamefont{Pryor}},\ and\ \bibinfo {author}
  {\bibfnamefont{M.~E.}\ \bibnamefont{Flatt\'e}},\ }%
  \bibfield{journal}{%
  \bibinfo {journal} {Appl. Phys. Lett.}\ }%
  \textbf{\bibinfo {volume} {92}},\ \bibinfo {pages} {222502} (\bibinfo {year}
  {2008})%
  \bibAnnoteFile{NoStop}{Pingenot2008}%
\bibitem{Andlauer2009}%
  \BibitemOpen
  \bibfield{author}{%
  \bibinfo {author} {\bibfnamefont{T.}~\bibnamefont{Andlauer}}\ and\ \bibinfo
  {author} {\bibfnamefont{P.}~\bibnamefont{Vogl}},\ }%
  \bibfield{journal}{%
  \bibinfo {journal} {Phys. Rev. B}\ }%
  \textbf{\bibinfo {volume} {79}},\ \bibinfo {pages} {045307} (\bibinfo {year}
  {2009})%
  \bibAnnoteFile{NoStop}{Andlauer2009}%
\bibitem{Salis2001}%
  \BibitemOpen
  \bibfield{author}{%
  \bibinfo {author} {\bibfnamefont{G.}~\bibnamefont{Salis}}, \bibinfo {author}
  {\bibfnamefont{Y.}~\bibnamefont{Kato}}, \bibinfo {author}
  {\bibfnamefont{K.}~\bibnamefont{Ensslin}}, \bibinfo {author}
  {\bibfnamefont{D.~C.}\ \bibnamefont{Driscoll}}, \bibinfo {author}
  {\bibfnamefont{A.~C.}\ \bibnamefont{Gossard}},\ and\ \bibinfo {author}
  {\bibfnamefont{D.~D.}\ \bibnamefont{Awschalom}},\ }%
  \bibfield{journal}{%
  \bibinfo {journal} {Nature}\ }%
  \textbf{\bibinfo {volume} {414}},\ \bibinfo {pages} {619} (\bibinfo {year}
  {2001})%
  \bibAnnoteFile{NoStop}{Salis2001}%
\bibitem{Doty2006}%
  \BibitemOpen
  \bibfield{author}{%
  \bibinfo {author} {\bibfnamefont{M.~F.}\ \bibnamefont{Doty}}, \bibinfo
  {author} {\bibfnamefont{M.}~\bibnamefont{Scheibner}}, \bibinfo {author}
  {\bibfnamefont{I.}~\bibnamefont{Ponomarev}}, \bibinfo {author}
  {\bibfnamefont{E.~A.}\ \bibnamefont{Stinaff}}, \bibinfo {author}
  {\bibfnamefont{A.~S.}\ \bibnamefont{Bracker}}, \bibinfo {author}
  {\bibfnamefont{V.~L.}\ \bibnamefont{Korenev}}, \bibinfo {author}
  {\bibfnamefont{T.~L.}\ \bibnamefont{Reinecke}},\ and\ \bibinfo {author}
  {\bibfnamefont{D.}~\bibnamefont{Gammon}},\ }%
  \bibfield{journal}{%
  \bibinfo {journal} {Phys. Rev. Lett.}\ }%
  \textbf{\bibinfo {volume} {97}},\ \bibinfo {pages} {197202} (\bibinfo {year}
  {2006})%
  \bibAnnoteFile{NoStop}{Doty2006}%
\bibitem{Nakaoka2007}%
  \BibitemOpen
  \bibfield{author}{%
  \bibinfo {author} {\bibfnamefont{T.}~\bibnamefont{Nakaoka}}, \bibinfo
  {author} {\bibfnamefont{S.}~\bibnamefont{Tarucha}},\ and\ \bibinfo {author}
  {\bibfnamefont{Y.}~\bibnamefont{Arakawa}},\ }%
  \bibfield{journal}{%
  \bibinfo {journal} {Phys. Rev. B}\ }%
  \textbf{\bibinfo {volume} {76}},\ \bibinfo {pages} {041301(R)} (\bibinfo
  {year} {2007})%
  \bibAnnoteFile{NoStop}{Nakaoka2007}%
\bibitem{Klotz2010}%
  \BibitemOpen
  \bibfield{author}{%
  \bibinfo {author} {\bibfnamefont{F.}~\bibnamefont{Klotz}}, \bibinfo {author}
  {\bibfnamefont{V.}~\bibnamefont{Jovanov}}, \bibinfo {author}
  {\bibfnamefont{J.}~\bibnamefont{Kierig}}, \bibinfo {author}
  {\bibfnamefont{E.~C.}\ \bibnamefont{Clark}}, \bibinfo {author}
  {\bibfnamefont{D.}~\bibnamefont{Rudolph}}, \bibinfo {author}
  {\bibfnamefont{D.}~\bibnamefont{Heiss}}, \bibinfo {author}
  {\bibfnamefont{M.}~\bibnamefont{Bichler}}, \bibinfo {author}
  {\bibfnamefont{G.}~\bibnamefont{Abstreiter}}, \bibinfo {author}
  {\bibfnamefont{M.~S.}\ \bibnamefont{Brandt}},\ and\ \bibinfo {author}
  {\bibfnamefont{J.~J.}\ \bibnamefont{Finley}},\ }%
  \bibfield{journal}{%
  \bibinfo {journal} {Appl. Phys. Lett.}\ }%
  \textbf{\bibinfo {volume} {96}},\ \bibinfo {pages} {053113} (\bibinfo {year}
  {2010})%
  \bibAnnoteFile{NoStop}{Klotz2010}%
\bibitem{pryor:2006}%
  \BibitemOpen
  \bibfield{author}{%
  \bibinfo {author} {\bibfnamefont{C.}~\bibnamefont{Pryor}}\ and\ \bibinfo
  {author} {\bibfnamefont{M.}~\bibnamefont{Flatte}},\ }%
  \bibfield{journal}{%
  \bibinfo {journal} {Phys. Rev. Lett.}\ }%
  \textbf{\bibinfo {volume} {96}},\ \bibinfo {pages} {026804} (\bibinfo {year}
  {2006})%
  \bibAnnoteFile{NoStop}{pryor:2006}%
\bibitem{Heyn2003a}%
  \BibitemOpen
  \bibfield{author}{%
  \bibinfo {author} {\bibfnamefont{C.}~\bibnamefont{Heyn}}\ and\ \bibinfo
  {author} {\bibfnamefont{W.}~\bibnamefont{Hansen}},\ }%
  \bibfield{journal}{%
  \bibinfo {journal} {Journal of Crystal Growth}\ }%
  \textbf{\bibinfo {volume} {251}},\ \bibinfo {pages} {140} (\bibinfo {year}
  {2003})%
  \bibAnnoteFile{NoStop}{Heyn2003a}%
\bibitem{Heyn2003b}%
  \BibitemOpen
  \bibfield{author}{%
  \bibinfo {author} {\bibfnamefont{C.}~\bibnamefont{Heyn}}\ and\ \bibinfo
  {author} {\bibfnamefont{W.}~\bibnamefont{Hansen}},\ }%
  \bibfield{journal}{%
  \bibinfo {journal} {Journal of Crystal Growth}\ }%
  \textbf{\bibinfo {volume} {251}},\ \bibinfo {pages} {218} (\bibinfo {year}
  {2003})%
  \bibAnnoteFile{NoStop}{Heyn2003b}%
\bibitem{Keizer2010}%
  \BibitemOpen
  \bibfield{author}{%
  \bibinfo {author} {\bibfnamefont{J.~G.}\ \bibnamefont{Keizer}}, \bibinfo
  {author} {\bibfnamefont{E.~C.}\ \bibnamefont{Clark}}, \bibinfo {author}
  {\bibfnamefont{M.}~\bibnamefont{Bichler}}, \bibinfo {author}
  {\bibfnamefont{G.}~\bibnamefont{Abstreiter}}, \bibinfo {author}
  {\bibfnamefont{J.~J.}\ \bibnamefont{Finley}},\ and\ \bibinfo {author}
  {\bibfnamefont{P.~M.}\ \bibnamefont{Koenraad}},\ }%
  \bibfield{journal}{%
  \bibinfo {journal} {Nanotechnology}\ }%
  \textbf{\bibinfo {volume} {21}},\ \bibinfo {pages} {215705} (\bibinfo {year}
  {2010})%
  \bibAnnoteFile{NoStop}{Keizer2010}%
\bibitem{andlauer:2008}%
  \BibitemOpen
  \bibfield{author}{%
  \bibinfo {author} {\bibfnamefont{T.}~\bibnamefont{Andlauer}}, \bibinfo
  {author} {\bibfnamefont{R.}~\bibnamefont{Morschl}},\ and\ \bibinfo {author}
  {\bibfnamefont{P.}~\bibnamefont{Vogl}},\ }%
  \bibfield{journal}{%
  \bibinfo {journal} {Phys. Rev. B}\ }%
  \textbf{\bibinfo {volume} {78}},\ \bibinfo {pages} {075317} (\bibinfo {year}
  {2008})%
  \bibAnnoteFile{NoStop}{andlauer:2008}%
\bibitem{wilson:1974}%
  \BibitemOpen
  \bibfield{author}{%
  \bibinfo {author} {\bibfnamefont{K.~G.}\ \bibnamefont{Wilson}},\ }%
  \bibfield{journal}{%
  \bibinfo {journal} {Phys. Rev. D}\ }%
  \textbf{\bibinfo {volume} {10}},\ \bibinfo {pages} {2445} (\bibinfo {year}
  {1974})%
  \bibAnnoteFile{NoStop}{wilson:1974}%
\bibitem{burt:1999}%
  \BibitemOpen
  \bibfield{author}{%
  \bibinfo {author} {\bibfnamefont{M.~G.}\ \bibnamefont{Burt}},\ }%
  \bibfield{journal}{%
  \bibinfo {journal} {J. Phys.: Condens. Matter}\ }%
  \textbf{\bibinfo {volume} {11}},\ \bibinfo {pages} {R53} (\bibinfo {year}
  {1999})%
  \bibAnnoteFile{NoStop}{burt:1999}%
\bibitem{stier:1999}%
  \BibitemOpen
  \bibfield{author}{%
  \bibinfo {author} {\bibfnamefont{O.}~\bibnamefont{Stier}}, \bibinfo {author}
  {\bibfnamefont{M.}~\bibnamefont{Grundmann}},\ and\ \bibinfo {author}
  {\bibfnamefont{D.}~\bibnamefont{Bimberg}},\ }%
  \bibfield{journal}{%
  \bibinfo {journal} {Phys. Rev. B}\ }%
  \textbf{\bibinfo {volume} {59}},\ \bibinfo {pages} {5688} (\bibinfo {year}
  {1999})%
  \bibAnnoteFile{NoStop}{stier:1999}%
\bibitem{trebin:1979}%
  \BibitemOpen
  \bibfield{author}{%
  \bibinfo {author} {\bibfnamefont{H.}~\bibnamefont{Trebin}}, \bibinfo {author}
  {\bibfnamefont{U.}~\bibnamefont{R\"ossler}},\ and\ \bibinfo {author}
  {\bibfnamefont{R.}~\bibnamefont{Ranvaud}},\ }%
  \bibfield{journal}{%
  \bibinfo {journal} {Phys. Rev. B}\ }%
  \textbf{\bibinfo {volume} {20}},\ \bibinfo {pages} {686} (\bibinfo {year}
  {1979})%
  \bibAnnoteFile{NoStop}{trebin:1979}%
\bibitem{offermans:2005}%
  \BibitemOpen
  \bibfield{author}{%
  \bibinfo {author} {\bibfnamefont{P.}~\bibnamefont{Offermans}}, \bibinfo
  {author} {\bibfnamefont{P.~M.}\ \bibnamefont{Koenraad}}, \bibinfo {author}
  {\bibfnamefont{J.~H.}\ \bibnamefont{Wolter}}, \bibinfo {author}
  {\bibfnamefont{K.}~\bibnamefont{Pierz}}, \bibinfo {author}
  {\bibfnamefont{M.}~\bibnamefont{Roy}},\ and\ \bibinfo {author}
  {\bibfnamefont{P.~A.}\ \bibnamefont{Maksym}},\ }%
  \bibfield{journal}{%
  \bibinfo {journal} {Phys. Rev. B}\ }%
  \textbf{\bibinfo {volume} {72}},\ \bibinfo {pages} {165332} (\bibinfo {year}
  {2005})%
  \bibAnnoteFile{NoStop}{offermans:2005}%
\bibitem{migliorato:2002}%
  \BibitemOpen
  \bibfield{author}{%
  \bibinfo {author} {\bibfnamefont{M.~A.}\ \bibnamefont{Migliorato}}, \bibinfo
  {author} {\bibfnamefont{A.~G.}\ \bibnamefont{Cullis}}, \bibinfo {author}
  {\bibfnamefont{M.}~\bibnamefont{Fearn}},\ and\ \bibinfo {author}
  {\bibfnamefont{J.~H.}\ \bibnamefont{Jefferson}},\ }%
  \bibfield{journal}{%
  \bibinfo {journal} {Phys. Rev. B}\ }%
  \textbf{\bibinfo {volume} {65}},\ \bibinfo {pages} {115316} (\bibinfo {year}
  {2002})%
  \bibAnnoteFile{NoStop}{migliorato:2002}%
\bibitem{keizer:2010}%
  \BibitemOpen
  \bibfield{author}{%
  \bibinfo {author} {\bibfnamefont{J.}~\bibnamefont{Keizer}}, \bibinfo {author}
  {\bibfnamefont{E.}~\bibnamefont{Clark}}, \bibinfo {author}
  {\bibfnamefont{M.}~\bibnamefont{Bichler}}, \bibinfo {author}
  {\bibfnamefont{G.}~\bibnamefont{Abstreiter}}, \bibinfo {author}
  {\bibfnamefont{J.}~\bibnamefont{Finley}},\ and\ \bibinfo {author}
  {\bibfnamefont{P.}~\bibnamefont{Koenraad}},\ }%
  \bibfield{journal}{%
  \bibinfo {journal} {IOP Nanotechnology}\ }%
  \textbf{\bibinfo {volume} {21}},\ \bibinfo {pages} {215705} (\bibinfo {year}
  {2010})%
  \bibAnnoteFile{NoStop}{keizer:2010}%
\bibitem{nakaoka:2004}%
  \BibitemOpen
  \bibfield{author}{%
  \bibinfo {author} {\bibfnamefont{T.}~\bibnamefont{Nakaoka}}, \bibinfo
  {author} {\bibfnamefont{T.}~\bibnamefont{Saito}}, \bibinfo {author}
  {\bibfnamefont{J.}~\bibnamefont{Tatebayashi}},\ and\ \bibinfo {author}
  {\bibfnamefont{Y.}~\bibnamefont{Arakawa}},\ }%
  \bibfield{journal}{%
  \bibinfo {journal} {Phys. Rev. B}\ }%
  \textbf{\bibinfo {volume} {70}},\ \bibinfo {pages} {235337} (\bibinfo {year}
  {2004})%
  \bibAnnoteFile{NoStop}{nakaoka:2004}%
\bibitem{nakaoka:2005}%
  \BibitemOpen
  \bibfield{author}{%
  \bibinfo {author} {\bibfnamefont{T.}~\bibnamefont{Nakaoka}}, \bibinfo
  {author} {\bibfnamefont{T.}~\bibnamefont{Saito}}, \bibinfo {author}
  {\bibfnamefont{J.}~\bibnamefont{Tatebayashi}}, \bibinfo {author}
  {\bibfnamefont{S.}~\bibnamefont{Hirose}}, \bibinfo {author}
  {\bibfnamefont{T.}~\bibnamefont{Usuki}}, \bibinfo {author}
  {\bibfnamefont{N.}~\bibnamefont{Yokoyama}},\ and\ \bibinfo {author}
  {\bibfnamefont{Y.}~\bibnamefont{Arakawa}},\ }%
  \bibfield{journal}{%
  \bibinfo {journal} {Phys. Rev. B}\ }%
  \textbf{\bibinfo {volume} {71}},\ \bibinfo {pages} {205301} (\bibinfo {year}
  {2005})%
  \bibAnnoteFile{NoStop}{nakaoka:2005}%
\bibitem{roth:1959}%
  \BibitemOpen
  \bibfield{author}{%
  \bibinfo {author} {\bibfnamefont{L.}~\bibnamefont{Roth}}, \bibinfo {author}
  {\bibfnamefont{B.}~\bibnamefont{Lax}},\ and\ \bibinfo {author}
  {\bibfnamefont{S.}~\bibnamefont{Zwerdling}},\ }%
  \bibfield{journal}{%
  \bibinfo {journal} {Phys. Rev.}\ }%
  \textbf{\bibinfo {volume} {114}},\ \bibinfo {pages} {90} (\bibinfo {year}
  {1959})%
  \bibAnnoteFile{NoStop}{roth:1959}%
\bibitem{kiselev:1998}%
  \BibitemOpen
  \bibfield{author}{%
  \bibinfo {author} {\bibfnamefont{A.}~\bibnamefont{Kiselev}}, \bibinfo
  {author} {\bibfnamefont{E.}~\bibnamefont{Ivchenko}},\ and\ \bibinfo {author}
  {\bibfnamefont{U.}~\bibnamefont{R\"ossler}},\ }%
  \bibfield{journal}{%
  \bibinfo {journal} {Phys. Rev. B}\ }%
  \textbf{\bibinfo {volume} {58}},\ \bibinfo {pages} {16 353} (\bibinfo {year}
  {1998})%
  \bibAnnoteFile{NoStop}{kiselev:1998}%
\bibitem{lommer:1985}%
  \BibitemOpen
  \bibfield{author}{%
  \bibinfo {author} {\bibfnamefont{G.}~\bibnamefont{Lommer}}, \bibinfo {author}
  {\bibfnamefont{F.}~\bibnamefont{Malcher}},\ and\ \bibinfo {author}
  {\bibfnamefont{U.}~\bibnamefont{R\"ossler}},\ }%
  \bibfield{journal}{%
  \bibinfo {journal} {Phys. Rev. B RC}\ }%
  \textbf{\bibinfo {volume} {32}},\ \bibinfo {pages} {6967} (\bibinfo {year}
  {1985})%
  \bibAnnoteFile{NoStop}{lommer:1985}%
\bibitem{foreman:2000}%
  \BibitemOpen
  \bibfield{author}{%
  \bibinfo {author} {\bibfnamefont{B.~A.}\ \bibnamefont{Foreman}},\ }%
  \bibfield{journal}{%
  \bibinfo {journal} {J. Phys.: Condens. Matter}\ }%
  \textbf{\bibinfo {volume} {12}},\ \bibinfo {pages} {R435} (\bibinfo {year}
  {2000})%
  \bibAnnoteFile{NoStop}{foreman:2000}%
\end{thebibliography}%

\end{document}